\documentclass[12pt,preprint]{aastex}

\newcommand{\kms}{\mbox{km~s$^{-1}$}}

\newcommand{\twodustalt}{{\bf 2-D}{\sc ust}~}

\shorttitle{Envelope of HD~56126}
\shortauthors{Meixner et al.}

\begin{document}

\title{The Molecular and Dust Envelope of  HD~56126}

\author{ M.~Meixner\altaffilmark{1,2}, 
 A.~Zalucha\altaffilmark{2},
T.~Ueta\altaffilmark{3,2},
D.~Fong\altaffilmark{4,2},  
 K.~Justtanont\altaffilmark{5}}

\altaffiltext{1}{Space Telescope Science Institute, 3700 San Martin Dr., 
 Baltimore, MD  21218}
\altaffiltext{2}{Astronomy Department, University of Illinois, 
 1002 W. Green St., Urbana, IL 61801}
\altaffiltext{3}{Royal Observatory of Belgium, Ringlaan 3, 1180 Brussels, Belgium}
\altaffiltext{4}{Harvard/CfA, SMA, Hawaii}
\altaffiltext{5}{SCFAB, Stockholm Observatory, Department of Astronomy, 106 91 Stockholm, Sweden}

\begin{abstract}
We present millimeter interferometry images of the CO J=1-0 line emission 
arising in the circumstellar envelope of 
HD 56126 (a.k.a.\ IRAS~07134+1005), which is one of the best studied 21-$\mu$m 
proto-planetary nebulae (PPNs). The CO emission extends from 1.2\arcsec\ to
7\arcsec\ in radius from the central star and appears consistent with a
simple expanding envelope, as expected for a post-AGB star.  The CO envelope is
very clumpy with no apparent fast wind to explain these microstructures that must
have arisen during the AGB mass-loss.  We quantitatively model the molecular envelope
using a radiative transfer code that we have modified for detached shells.  
Our best fit model reveals that two sequential winds created the circumstellar envelope of
HD 56126: an AGB wind that lasted 6500 years  with a mass-loss rate of 
$5.1\times 10^{-6}$ M$_\odot$ yr$^{-1}$ and a more intense superwind that lasted 840 years
with a mass-loss rate of $3\times 10^{-5}$ M$_\odot$ yr$^{-1}$ and that ended
the star's life on the AGB 1240 years ago. The total mass of this
envelope is 0.059 M$_\odot$ which indicates a lower limit progenitor mass for the system
of 0.66 M$_\odot$, quite reasonable for this  low-metallicity star which probably
resides in the thick disk of the Galaxy. Comparison with images
of the dust emission reveal a similar structure with the gas in the inner regions.
Using \twodustalt, we model the dust emission of this source so that the model is consistent 
with the CO emission model and find a total dust mass of $7.8\times 10^{-4} $ M$_\odot$, 
a superwind dust mass-loss rate of $ 1.9\times 10^{-7}$ M$_\odot$ yr$^{-1}$,
and an AGB dust mass-loss rate of $9.6\times 10^{-8}$ M$_\odot$ yr$^{-1}$.  
We derive an average gas-to-dust mass ratio of 75,
which is consistent with ISM values, but low for what most consider for carbon stars.  Our results
indicate that TiC nanocrystals are probably not the carrier of the 21-$\mu$m feature. 
\end{abstract}


\section{Introduction}

The proto-planetary nebula (PPN) stage of intermediate mass
(0.8--8.0 M$_\odot$) stellar evolution is a very
short lived ($\sim$1000 years) phase occurring after the asymptotic
giant branch (AGB) and before a planetary nebula (PN) is formed
\citep{kwok1993,vanwinckel2003}.  A sub-class of PPN are
characterized by an unidentified infrared feature at 21 $\mu$m
and have been dubbed the 21-$\mu$m PPNs \citep{kwok1989,volk1999}.
\citet{vonhelden2000} 
have proposed nanocrystals of titanium carbide (TiC) to be the carrier of 
the 21-$\mu$m feature.   However, in order to create TiC high
densities are required in the circumstellar environment and that
means high mass loss rates, on the order of $10^{-3}$ M$_\odot$ yr$^{-1}$.
In fact, \citet{vonhelden2000}  suggest that the entire circumstellar
envelope was created in a singular catastrophic mass loss event.

HD 56126 (a.k.a.\ IRAS~07134+1005) is one of the best 
studied 21-$\mu$m PPN. The central star is variable with a period of 36.8 days
indicating a mass of 0.6 M$_\odot$ \citep{barthes2000}.
Abundance analysis by \citet{vanwinckel2000} reveals a metal
poor star with less than solar Fe abundance, but with solar or greater than solar
abundances of C, N, O and s-process elements indicating the star experienced third dredge-up
when it was on the AGB.  Indeed, it is considered a carbon rich source and has evidence for
near-IR bands attributed to polycyclic aromatic hydrocarbons (PAHs) in addition to its
strong 21-$\mu$m feature \citep{volk1999}.  Optical HST images of HD 56126 reveal a bright
central star surrounded by a low surface brightness elliptical nebula \citep{ueta2000}. 
The many mid-IR images \citep{meixner1997,dayal1998,jura2000,kwok2002} 
of HD 56126 reveal a detached shell with two peaks aligned on an
east-west axis that has been interpreted by some as limb brightened peaks of an equatorial
density enhancement \citep{meixner1997,dayal1998}.  
Recent near-IR imaging polarimetry of HD 56126 reveal a thin
(both geometrically and optically), limb-brightened, and
well-structured shell with an equatorial density enhancement
\citep{ueta2004}.
Despite these many beautiful images of the dust there are no published images of
the circumstellar gas emission which has been detected in several CO transitions \citep{zuckerman1986,
bujarrabal1992,knapp1998,knapp2000} and in the CI 609 $\mu$m line \citep{knapp2000}.  All
current mass loss rate estimates are based on the dust emission with assumed 
gas-to-dust mass ratios.

In order to determine the structure  of the molecular envelope and to constrain  the gas mass-loss
rate history of HD 56126, we have pursued  imaging of the CO J=1-0 line.
Here we present the results of this new imaging study (sections 2 and 3).
A quantitative analysis of these images and previously published
higher CO transitions using a radiative transfer code is presented in
section 4.  In section 5, we derive the dust mass and mass-loss rate history in
a consistent fashion with the CO modeling. We discuss the implications of
our observations and models in section 6.  We summarize our conclusions
in section 7.

\section{BIMA Observations}

Using the Berkeley-Illinois-Maryland Association (BIMA\footnote{Operated
by the University of California, Berkeley, the University of
Illinois, and the University of Maryland, with support from
the National Science Foundation}) millimeter array \citep{welch1996},
we observed  HD 56126  in the CO J=1-0 transition at
115.2712 GHz in the A, B, C and D array configurations.   
We used a 7-point hexagonal mosaic with overlapping fields with offsets of 30\arcsec\ to
ensure uniform sensitivity to structure within the central 110\arcsec\ and 
to have a HPBW of 160\arcsec. The center of this mosaic was
R.A.(J2000) = 07$^h$ 16$^m$ 10.3$^s$ and  Dec. (2000) = +09\arcdeg 59\arcmin 48\arcsec, which
is 0\farcs63 east of the position of the central star, 
R.A.(J2000) = 07$^h$ 16$^m$ 10.26$^s$ and  Dec. (2000) = +09\arcdeg 59\arcmin 48\arcsec,
\citep{ueta2000}. The correlator was configured to cover a bandwidth of 50 MHz, resulting
in a velocity coverage of 128 \kms\, with a spectral resolution of 1 \kms.  
The final images contain 
just B, C, and D array configuration data, because no signal was
detected in A array and thus we do not discuss the A array further.
The dates of observation included were:  31 Mar.\ 2001, 29
Apr.\ 2001, 7 May 2001 (C array); 16 Jun.\ 2001, 7 Jul.\ 2001 (D array); and
2 Mar.\ 2002 (B array).  The phase calibrator for all arrays was quasar 0739+016, which was 
chosen from the BIMA phase/amplitude calibrator list.
The flux calibrator was Saturn for D array and Jupiter for C and B
array. The system temperature for the included observations ranged from 
300 K to 1000 K.  The {\it uv} coverage ranged from 2.3 to 85 k$\lambda$.  Using the MIRIAD software package
(Sault, Teuben, \& Wright 1995), we followed standard data calibration, imaging and 
deconvolution procedures.  Robust weighting with a robust parameter $= -1$ was used to
weight the visibility data for images, resulting in a beam size of 
3.5\arcsec $\times$ 2.9\arcsec. As a final step, we used an iterative self-calibration
procedure on the data to remove some residual phase errors.  We analyzed the data further using
standard routines in MIRIAD.
 
\section{BIMA Results}

Our BIMA observations (Figure \ref{cofig}; Table \ref{bimatab}) reveal the entire
molecular envelope down to the sensitivity limit listed in Table \ref{bimatab}.
In particular, they are not missing flux due to the lack of compact spacings because our
total flux of the CO J=1-0 line is consistent  with the single dish telescope 
measurement \citep{bujarrabal1992} within a 30\% absolute flux calibration error. 
We also made pure continuum maps using the lower sideband
data; however, finding no emission, we only report an upper limit in  Table \ref{bimatab}.

The BIMA channel maps reveal a clumpy, molecular envelope expanding away from the star 
at V$_{exp}= 10\pm 1$ \kms, which is measured as the half width at zero intensity level
(Fig.~\ref{cofig}).  The most blue- and red- shifted channels show a compact CO emission 
structure which increases
in size as one approaches the central channel maps as one expects from an expanding
envelope. Our systemic velocity,V$_{LSR} = 73\pm$1 \kms, is the  central velocity
of the CO J=1-0 line.  The CO is distributed around  the central star, marked by the cross, 
which is located $-$0.\arcsec 63 in RA from the map center.  
A number of clumps in the envelope gives it
the appearance of a non-spherical envelope with protrusions and clumps in all directions.
The distribution in spatial and velocity directions suggests an almost random distribution of
the protrusions and clumps and there appear to be no fast, collimated 
outflows to create these structures.  The total flux line profile is parabolic with no line
wings, as expected for simple expanding outflows (Fig.~\ref{cofig}). An azimuthal average 
of the CO emission at the velocity 73 \kms\ reveals an outer radius of 
$\sim$7\arcsec\ (Fig.~\ref{cofig}).

Figure~\ref{comidirhst} shows a comparison of  CO emission averaged over the
velocity range 66 to 78 \kms\  with the mid-infrared thermal
dust emission \citep{kwok2002}  and the dust-scattered light from the circumstellar 
shell \citep{ueta2000}.  We average over this velocity range because the
dust emission is a projection of all the emission along the line of sight.  
The warm dust emission is completely contained within
the larger CO emission. The combined B, C and D array data does not have sufficient
angular resolution to resolve the inner cavity. Thus,  we have made a
higher angular resolution (2.6\arcsec $\times$ 2.0\arcsec) map using
only B array data  that covers the velocity channels 66$-$78 km s$^{-1}$ (Fig.~\ref{comidirhst}).
Direct comparison of this B-array image with the mid-infrared images of the 
dust shell thermal emission  reveal a cavity quite similar 
in size and morphology in both the molecular gas and dust.
Based on this comparison, we conclude that the inner edges of
the dust and molecular gas envelopes are identical.
In particular, the central star is located in a clear depression at the center of both gas
and dust tracers  with an angular inner radius of $\sim 1.2$\arcsec.   
We  measure our inner radius as half of the 
distance between the two emission peaks to the east and west of the
central star. This approach results in a slightly larger value than
the $\sim$0.\arcsec 8 adopted by \cite{kwok2002}.  
Our model calculations, described below, support our value for the radius.

HD 56126 has been observed with single dish telescopes in the CO J=2-1 line
\citep{knapp1998}, the CO J=4-3 line \citep{knapp2000} and the [CI] 609 $\mu$m line 
\citep{knapp2000}.   The detection of the [CI] 609 $\mu$m line  emission is
particularly interesting because it indicates the photodissociation of molecules
in the envelope.  \cite{knapp2000} suggest that the [CI] arises from
the dissociation of CO due to shocks from a fast wind because the central star,
F5 Iab, is too cool to photodissociate the CO. However, our results indicate
that the CO is not dissociated because it has the same inner radius as the
dust emission.  Instead, we suggest that the [CI] is
created by the photodissociation of C$_2$H$_2$ which is expected to be
the next most abundant molecule to CO in carbon rich sources and which has been suggested as
the source for [CI] emission in AFGL 2688 \citep{fong2001}.  The  photodissociation
potential of C$_2$H$_2$, 6.2 eV, is much lower  than that of CO,
11 eV, and at a T$_{eff}$ = 7250 K approximately 1\% of HD56126's bolometric
luminosity has sufficient energy to dissociate   C$_2$H$_2$ making 
this avenue quite plausible  \citep{fong2001}. If photodissociation produces
[CI] emission, then photodissociation most likely influences the energetics of the molecular  
gas  and the CO emission in the circumstellar environment of HD 56126.

\section{CO Modeling \label{comodel}}

In order to constrain the gas mass loss history of HD 56126, we have modeled 
the CO J=1-0 BIMA  data cube, and the CO J=2-1 and J=4-3 line profiles of \cite{knapp1998}
and \cite{knapp2000}, respectively.    

\subsection{Model code}

In order to model this proto-planetary nebula,
we have adapted the radiative transfer code of 
\cite{justtanont1994} that was originally constructed for 
AGB star circumstellar envelopes which have molecular gas extending
from the photosphere to the outer circumstellar envelope.  For
proto-planetary nebulae, the mass loss ended some time ago creating
a detached shell, such as we observe in HD 56126.   
We have modified the possible density profiles to include a gap
at the center and multiple episodes of mass loss and have modified the
code to properly handle the gap.  The density, $\rm \rho_{H_2}(R)$, as
a function of radius, R,  follows the relations  

$$\rm   \rho_{H_2}(R)  = 10^{-20} ~~~~~~~~~~~~~ R < R_{in} $$
$$\rm   \rho_{H_2}(R)  =  {\dot{M} \over 4\pi R^2 V_{exp} }~~~~~~~~~~~~~ R_{in}  < R < R_{SW} $$
$$\rm   \rho_{H_2}(R)  =  F\times {\dot{M} \over 4\pi R^2 V_{exp} }~~~~~~~~~~~~~ R_{SW}  < R < R_{out} $$

\noindent where \.M is the mass loss rate, $\rm  V_{exp} $ is the expansion velocity, F is the factor by which
mass loss was higher or lower in the past, $\rm   R_{in}$ is the inner radius, $\rm  R_{SW}$ is the superwind
radius, and  $\rm  R_{out} $ is the outer radius of the envelope.                                   

The  gas temperature, $\rm   T_{gas}$, as a function of radius, R,  
was  also modified to allow for an inner gap and different power laws 
over the different mass loss episodes with sudden drops at the boundaries.

$$\rm   T_{gas}(R)  =  2.8 K ~~~~~~~~~~~~~ R < R_{in} $$
$$\rm   T_{gas}(R)  =  T_{in} ({R\over R_{in}})^{-\epsilon} ~~~~~~~~~~~~~ R_{in}  < R < R_{SW} $$
$$\rm   T_{gas}(R)  =  T_{in}\times F_T \times ({R_{SW}\over R_{in}})^{-\epsilon} \times  
({R\over R_{SW}})^{-\epsilon_2}  ~~~~~~~~~~~~~ R_{SW}  < R < R_{out} $$

\noindent The gas temperature, $\rm T_{in}$, 
at the inner radius, $\rm R_{in}$, drops with respect to radius 
as a power law with an exponent, $\epsilon$, until
the superwind radius, $\rm  R_{SW}$. At $\rm  R_{SW}$, the temperature can drop by a factor of
$\rm F_T$ and then continue to drop  with respect to radius 
as a power law with a different exponent, $\epsilon_2$.
Figure \ref{dentempfig}  shows  the temperature and density profiles that we adopted for
HD 56126 and that we discuss below.

\subsection{Model Parameters}

Table \ref{modelcotab}  lists the assumed and derived properties in the CO radiative transfer model.
Some of the assumed  properties, e.g.\ the distance,  are taken directly  from the literature. While others,
e.g.\ the CO/H$_2$ ratio,  are derived from information in the literature. The stellar radius, R$_*$, 
is used as the unit of scaling for the size of the dust shell and we derive this radius
from the luminosity, L$_*$, and effective temperature, T$_{eff}$, in our dust modeling.  
The inner and  outer radii are constrained by our BIMA
CO J=1-0 line images.  Using the CO J=1-0 line to constrain the model, and checking with
the CO J=4-3 and J=2-1 lines , we refine the values for the
expansion velocity, V$_{exp}$ = 10.5$\pm 0.5$ \kms,  and  systemic velocity,  V$_{LSR}$ = 73$\pm 1$  \kms. 

HD 56126 is considered a carbon rich source with a C/O $\sim$ 1.0 \citep{vanwinckel2000}.   
In the molecular gas, we  assume most of the carbon and oxygen is contained in CO. 
\cite{vanwinckel2000} derived a  photospheric carbon abundance of [C/H] = 0.1 which is defined 
with respect to solar abundances.  Converting this value to a number density of carbon with respect
to hydrogen by using the solar abundances of \cite{grevesse1989}, we find
a carbon number density, C/H, of  $4.6\times 10^{-4}$.  In the 
circumstellar molecular envelope, the H becomes H$_2$ and the C becomes CO. 
Thus the CO/H$_2$ $\sim 9.2\times 10^{-4}$ because the molecular hydrogen number density 
is now half that of the C which has become CO.  This value is on
the high side but within the $10^{-5}$--$10^{-3}$ range  
assumed for carbon rich sources by \cite{knapp1985}. Changes in the CO/H$_2$
will result in a linearly proportional change in the  final mass and mass-loss rate values.

The output parameters are model results constrained by the CO  observations. The CO emission depends upon the 
adopted temperature and density profiles (Fig.~\ref{dentempfig}).The mass-loss rates are directly related
to the density profile.  We ran over 80 models adjusting primarily the parameters that describe the temperature
and density profiles.  We started with the simplest of models,  single, constant mass loss rate  and single power law 
temperature. However,  finding it inadequate, we gradually  increased the model's complexity until the data
were adequately fit.

While simple power laws for the temperature profiles 
are appropriate for AGB stars, we find something different for this PPN which has
markedly different energetics than an AGB circumstellar envelope. 
The temperature profile could not be described by a single power law since this  would overestimate the CO J=1-0 if we
fit the CO J=4-3 line or underestimate the CO J=4-3 line if we fit the CO J=1-0 emission.     
The best fit temperature profile required distinct temperature laws for the superwind region and
the AGB wind region (Fig.~\ref{dentempfig}).  In the superwind region, close to the central star,
the gas has a temperature constrained to be 150 K, the inner temperature of the dust shell, 
because we assume the gas and dust are in thermal equilibrium.
The gas temperature then drops steeply with a  power law of $-$2.0.  
At the superwind radius, R$_{SW}$, the gas temperature suddenly drops by
a factor of 0.43 (F$_T$). In the following AGB wind region, the gas temperature
is low, 20 K,  and rather flat, with a power law of $-$0.43.    
These two distinct temperature profiles crudely describe two 
different physical processes that dominate the energetics of the two regions.

In the superwind region, the gas temperature is governed by the photodissociation region (PDR) 
created by the central star.   The presence of [CI] 609 $\mu$m emission
argues for photodissociation in the inner edge of the envelope. \cite{fong2001} presented
a PDR model for carbon rich proto-planetary nebulae and convincingly
explained the atomic fine structure line emission of PPN as arising from a PDR.  
The shape of our temperature profile (Fig.~\ref{dentempfig}) is a crude approximation
to the shape of the temperature profile predicted by Fong et al.'s PDR model (their Fig.~7) which
includes chemistry and balances the heating and cooling of the gas.
At the inner edge, the temperature is the highest because the photo-electric heating 
is most effective nearest to the star.  Collisional heating of the gas by warm dust grains
is a secondary heating mechanism for the gas. The temperature decreases radially because the
photo-electric heating  and the dust temperature both decrease with radius.

In the AGB wind region,  the sudden decrease in temperature in our model may be explained by 
the sudden decrease in density (see below) which would reduce the effectiveness of either  heating
process.
The outer AGB wind region  temperature is governed by a PDR created by the ISRF.
The flatness of the temperature law in 
the AGB wind region crudely approximates the flatter or perhaps rising gas temperature expected by
photoelectric heating by the interstellar radiation field (ISRF).  Evidence for this
effect has been found in the detailed models for the carbon star, IRC+10216 \citep[e.g.\ by][]{huggins1988},
and for PPN, AFGL 618 \citep{meixner1998}.

The density profile could not be defined by a single, constant mass-loss rate (Fig.~\ref{dentempfig}).  
Instead we adopted the two mass-loss epochs approach, i.e.\ superwind and AGB wind,
that have been required  in our previous CO modeling efforts of evolved stars \citep{meixner1998,fong2002}.  
In the case of HD~56126, the CO J=1-0 radial intensity
profile (Fig.~\ref{cofig}) required a sudden drop in intensity at a radius of   7.2$\times 10^{16}$ cm (2\arcsec),
which we define as the superwind radius, R$_{SW}$. The superwind mass-loss rate interior to this radius is 
substantially higher than the AGB mass-loss rate exterior to this radius. Both the superwind mass-loss rate and 
AGB mass-loss rate are constant for their duration.  We note that other PPN and young-PN 
that have been modeled in CO have also required a superwind and AGB wind to fit the data \citep{meixner1998}.

We derive the time scales by dividing the size of the regions  by the expansion velocity, V$_{exp}$.
The dynamical age, t$_{dyn}$, is the time since the star left the AGB and corresponds to the inner
radius, R$_{in}$, of the gas and dust shell.  The SW duration is the time scale during which the star lost
mass at the superwind mass-loss rate and corresponds to R$_{SW}$--R$_{in}$.  Finally, the AGB duration is the
time scale during which the star lost mass at the AGB mass-loss rate and corresponds to R$_{AGB}$--R$_{SW}$.

\subsection{Model results}

In order to compare the model results with the BIMA data cubes,  we have made a data cube
of the model CO J=1-0 line emission, read it into MIRIAD and convolved it with the same clean
beam parameters as the BIMA data   as previously done in \cite{fong2002}.  For comparison with the
CO J=2-1 and J=4-3 line transitions that were observed with single dish telescopes, we created a 
line profile with
the line flux spatially  weighted by a single dish beam profile comparable to the observed beam. We 
subtracted the systemic velocity from the observed CO J=2-1 and J=4-3 line profiles so we compare them
on the velocity scale relative to a systemic velocity.  
The results of the best fit model are compared with the CO data in 
Figures \ref{cofig} and \ref{linecofig}.  

The model fits, within the observed errors, the average 
properties of the  CO J=1-0  line emission. In particular the model's
approximate size of the CO emission in the channel maps and the
azimuthally averaged radial profile of the systemic velocity, 73 km s$^{-1}$,
are in good agreement.  The total flux line profile, which sums all the flux
in each channel map, also has good agreement between the data and model.
However, the clumps in the molecular gas surrounding HD 56126 are
responsible for most of the disagreement between the data and observations.
Modeling these clumps is beyond the scope of this work.  

The model CO J=4-3 line profile agrees reasonably with the data except that it
is somewhat wider in velocity at the half maximum flux points.
However, the model CO J=2-1 line profile 
is  higher than the data by $\sim$50\%.  We tried many iterations on the 
modeling to improve the agreement; however, forcing the model to fit
the CO J=2-1 observations causes the model to be discrepant with the J=1-0 
line data or the J=4-3 line data or both. This discrepancy between our
model CO J=2-1 line profile may suggest a more intricate temperature
profile is needed, because our profile only mimics a photodissociation
region profile, but it doesn't result from an energy balance solution.
Alternatively, the discrepancy could be due to an observational pointing
error on the source.  Mapped images of HD 56126 in the higher CO transitions,
such as are now possible with submillimeter arrays,
will be necessary to improve the modeling or to check the observations.

\section{Dust Model}

We use \twodustalt$\!\!$,\footnote{The \protect{\twodustalt}code is a general purpose
2-D dust radiative transfer code and is publicly available
(See http://www003.upp.so-net.ne.jp/ueta/research/2dust/ or
http://homepage.oma.be/ueta/research/2dust/ for details).} 
which is described in detail by  \cite{ueta2003}, 
to model the dust radiative transfer in this source
and to derive dust mass-loss rate parameters.   Our main goal is to derive a dust shell
model  that is consistent
with the CO observations  and reproduces the images of dust emission and scattering  
and the spectral energy distribution (SED) in an approximate manner.  Because  our CO
model is restricted to spherical symmetry while \twodustalt
allows an axial symmetry,  we define consistency between the two models as having the
same sizes for R$_{in}$, R$_{SW}$ and R$_{out}$ and 
the same mass loss recipe ($\dot{M}_{\rm AGB} < \dot{M}_{\rm SW}$); 
however, we do not expect identical morphologies.

Table \ref{modeldusttab} lists the input and output parameters.  
In addition, we adopt the dust optical constants for three kinds of
amorphous carbon species as described by \cite{zubko1996} and assumed
the power-law plus an exponential fall-off size distribution
\citep{kim94}.
We correct the model SED for interstellar extinction as described
by \cite{mathis1990}.
Input parameters are adopted whenever possible from independent measurements. 
The output parameters, which are results from the best-fit model as constrained by
the observed SED and  images, quantify the geometry of the mass loss, i.e. density
function parameters A, B, C, D, E, and F, \citep[cf.][]{ueta2003}; the equatorial
opacity, $\tau_{eq} (10.3)$; temperature, T$_{in}$(dust); mass, M$_{dust}$; 
and mass-loss rates, \.M$_{SW}$(dust) and \.M$_{AGB}$(dust), of the dust.

Figure \ref{moddustfig} compares  model images of dust
scattered light  (410 nm and 1.2 $\mu$m) and  thermal dust emission (10.3
and 18.0 $\mu$m).  The comparison between model and observed images over a wide
spectral range from optical to mid-IR is quite reasonable and
an improvement over our initial work in \cite{meixner1997}.
The improvement is due to a more detailed density function, 
defined in \citep{ueta2003}, and a more realistic description
of the optical properties of dust grains, which allows the
distribution of grain size and the use of multiple grain
species.   Figure  \ref{sedmodfig} shows that the  model SED and the observed
photometry and spectroscopy are in reasonably good agreement. The SED fit
is comparable to that in \cite{meixner1997} although much better constrained by
the data.  
The SED fit in the IR part of the spectrum is not as good as that by
\cite{hony2003}, who focused on the dust composition of the source and included
MgS and TiC in addition to the amorphous carbon in order to fit the 30 $\mu$m and
21-$\mu$m features in the spectrum: we used only amorphous carbon.  
Despite our differences in dust composition,
our final total dust mass, $7.8\times 10^{-4} $ M$_\odot$, is within $\sim$10\%
of \cite{hony2003}, because the amorphous carbon dominates their dust mass.
Our dust opacity along the equatorial plane, $\tau{eq}$, and
dust temperature at the inner radius on the equatorial plane,
T$_{dust}$, are comparable to values by \cite{meixner1997}. 
When corrected for the differences in distance and dust-to-gas mass ratio, 
our superwind dust mass-loss rate, \.M$_{SW}$(dust)$= 1.9\times 10^{-7}$ M$_\odot$ yr$^{-1}$,
is slightly higher (110\%) than that derived by \cite{meixner1997}  and our AGB dust mass-loss rate,
\.M$_{AGB}$(dust)$= 9.6\times 10^{-8}$ M$_\odot$ yr$^{-1}$, is substantially higher (factor of 7).
Nevertheless, our dust mass is significantally
smaller (30\%)  than that of  \cite{meixner1997} because the size of our
dust shell is constrained to be much smaller by the CO observations.

\section{Discussion}

\subsection{Circumstellar Envelope}

The mass-loss rates derived from the CO modeling, 
\.M$_{SW} \sim 3\times 10^{-5}$ M$_\odot$ yr$^{-1}$ and
\.M$_{AGB} \sim 5.1\times 10^{-6}$ M$_\odot$ yr$^{-1}$, are 
compatible with mass-loss rates measured for AGB stars, but
two orders of magnitude lower than the highest mass-loss rate measured in 
superwinds of proto-planetary nebulae, e.g. the Egg nebula \citep{skinner1997}.
Our values are consistent with that of \citet{jura2000} and of
\citet{meixner1997}, when scaled for distance. However, it is lower
by a factor of 10 from \citet{hony2003}  because they assumed a
more compact shell than we measure in the CO and because they assume a 
gas-to-dust ratio (222) that is three times higher than we derive.
 
The total mass of the observed molecular envelope is 0.059 M$_\odot$ which
is a lower limit to the total mass lost by the star because the outer
radius of the CO emission is limited by photodissociation from the
interstellar radiation field.  Adding this circumstellar mass to the
central star mass of 0.6 M$_\odot$ \citep{barthes2000} 
results in a lower limit to the progenitor mass
of HD 56126 of 0.66 M$_\odot$ which is consistent with the low-metallicity
of the central star that is probably a member of the thick disk of our
Galaxy. HD 56126 is clearly on the lower mass end of intermediate mass stars that may pass
through the planetary nebula phase.  

\subsection{Gas-to-Dust Mass ratio}

We can derive an average gas-to-dust mass ratio by dividing the total gas mass by
the total dust mass as determined by our independent, but consistent models
of the gas and dust shells.  We find a gas-to-dust mass ratio of 75  which
is close to values typically assumed for the interstellar medium  but lower
than what has been assumed for carbon stars, 222 \citep{jura1986}. 
If we were to derive the gas-to-dust mass ratio separately for the superwind
and AGB wind, we find a higher gas-to-dust mass ratio of 160 for the superwind
compared to a value of 50 for the AGB wind.  This difference may suggest
that the AGB wind was more dust rich than the superwind.  However, because
the CO model assumes spherical symmetry and the dust model assumes a toroidal
symmetry, the differences between AGB and super winds are tentative, at best.

\subsection{TiC as the carrier of the 21-\boldmath{$\mu$}m feature}

Two results from our study show a lack of support for TiC as the interpretation
for the 21-$\mu$m feature.  Firstly,  the size of the circumstellar shell is
at least twice as large as claimed by \cite{hony2003}, thus the mass was not lost
all at once in a catastrophic event.  Secondly,  the gas mass-loss
rates, \.M$_{SW} \sim 3\times 10^{-5}$ M$_\odot$ yr$^{-1}$ and
\.M$_{AGB} \sim 5.1\times 10^{-6}$ M$_\odot$ yr$^{-1}$, are one to two orders of magnitude
smaller than required by \cite{vonhelden2000} to create a high enough density environment
in which to produce TiC.

Our results support two other works which have cast doubt on TiC as the carrier
of the 21-$\mu$m feature. \cite{li2003} rules it out on the basis of the Kramers-Kronig
physical principle.  \cite{chigai2003} claim that  TiC is implausible as a carrier 
because it would require a Ti/Si abundance ratio 5 times that of solar to create the
strength of the 21-$\mu$m compared to the 11.3 $\mu$m SiC feature and, the most likely
case of TiC-core-carbonaceous-mantle grains exhibit only a weak 21-$\mu$m feature.

However, despite these failings of TiC nanocrystals to explain the 21-$\mu$m feature, the proposal
by \citet{vonhelden2000} has opened up a set of possible candidates:  nanocrystals containing
carbon.  Recently two proposals with nanocrystals claim to produce a 21-$\mu$m feature.
\citet{speck2004} show some laboratory spectra of SiC nanocrystals which show a 
21-$\mu$m feature. \citet{jones2004} have proposed nano-diamonds to explain the 21-$\mu$m feature.
Nano-crystals make some sense in that this feature is found primarily in proto-planetary
nebulae where the rapid changes in the central star wreck havoc on its circumstellar environment
which could cause the existence of very small grains.
In any case, the 21-$\mu$m feature carrier remains a mystery.

\section{Conclusions}

\begin{enumerate}

\item The CO emission reveals a resolved, but clumpy molecular envelope that extends
to $\sim$7\arcsec\ in radius.

\item Comparison of the CO images and mid-IR images of dust emission shows
identical inner radii demonstrating that the CO is not photodissociated; however,
C$_2$H$_2$ probably is dissociated and responsible for the [CI] 609 $\mu$m line emission.

\item The gas mass-loss rate was $3\times 10^{-5}$ M$_\odot$ yr$^{-1}$ during the
superwind phase and $5.1\times 10^{-6}$ M$_\odot$ yr$^{-1}$ during the AGB phase.
The dust mass-loss rate was $ 1.9\times 10^{-7}$ M$_\odot$ yr$^{-1}$ during the
superwind phase and $9.6\times 10^{-8}$ M$_\odot$ yr$^{-1}$ during the AGB phase.

\item The total gas mass of the circumstellar shell is 0.059 M$_\odot$ and dust mass is
$7.8\times 10^{-4} $ M$_\odot$, which is consistent
with HD 56126 being a lower mass proto-planetary nebula.

\item The average gas-to-dust mass ratio is 75, comparable to that typically assumed for the
ISM.

\item  TiC nanocrystals are implausible carriers of the 21-$\mu$m features.
The 21-$\mu$m feature's carrier remains a mystery.

\end{enumerate}

\acknowledgments

We are grateful for the financial support of several funding agencies.
Meixner and Zalucha were partially supported by NSF CAREER
grant AST 97-33697. Meixner, Ueta and Zalucha were partially supported by
NASA/STScI grant GO-9377.05-A.  Fong and Meixner were partially supported
by NSF grant AST 99-81546 and the Laboratory of Astronomical Imaging
at the University of Illinois. Meixner was partially supported by STScI/DDRF
grant D0001.82301. Ueta was partially supported by IUAP P5/36 financed
by the OSTC of the Belgian Federal State. This research has made use of the SIMBAD database, 
operated at CDS, Strasbourg,France.  Conversations with W. B. Latter on 
PDRs in circumstellar envelopes and with S. Hony on the circumstellar dust 
in HD 56126 were helpful in forming this paper.  Comments by the referee 
were useful in improving the paper.

\begin{deluxetable}{lc}
\tablewidth{0pt}
\tablecaption{BIMA Observations of  HD~56126\label{bimatab}}
\tablehead{
\colhead{Parameter} &\colhead{Value}}% \\
\startdata
RA (2000)\tablenotemark{1} & 07:16:10.3  \\
DEC (2000)\tablenotemark{1} & 09:59:48.0 \\
Beam size & 3.5\arcsec $\times$ 2.9\arcsec  \\
PA  & 16.4\arcdeg \\
Noise per chan  & 0.15 Jy Beam$^{-1}$  \\
BIMA CO Flux & 140 Jy km s$^{-1}$ \\
Single Dish Flux\tablenotemark{2} & 110 Jy km s$^{-1}$ \\
Chan width & 2 km s$^{-1}$ \\
V$_{LSR}$ & 73$\pm 1$ km s$^{-1}$   \\
V$_{exp}$ & 10$\pm 1$ km s$^{-1}$ \\
$\theta_{in}$ & 1\arcsec  \\
$\theta_{out}$ & 7\arcsec \\
continuum (2.6 mm, 3$\sigma$) & $< 8$ mJy \\ 
\enddata
\tablenotetext{1}{The star is located 0\farcs63 west of the map center at
RA (2000) = 07:16:10.26 and DEC (2000) = 09:59:48.0}
\tablenotetext{2}{\cite{bujarrabal1992} }
\end{deluxetable}

\begin{deluxetable}{lcc}
\tablewidth{0pt}
\tablecaption{Model Parameters  of HD~56126:  CO emission\label{modelcotab}}
\tablehead{
\colhead{Parameter} &\colhead{Value} &\colhead{Ref.}}%\\
\startdata
\textit{Input:} \\
D  & 2.4 kpc & 1 \\
R$_*$ & 3.45$\times 10^{12}$ cm & 1,2,3 \\
C/O & 1.0 & 2 \\
CO/H$_2$ & $9.2\times 10^{-4}$ & 2,3  \\
R$_{in}$ & 4.3$\times 10^{16}$ cm & 3 \\
R$_{out}$ & 3.0$\times 10^{17}$ cm & 3 \\
V$_{LSR}$ & 73$\pm 1$ km s$^{-1}$ & 3   \\
V$_{exp}$ & 10.5$\pm 0.5$ km s$^{-1}$  & 3\\
\\
\textit{Output:}  \\
R$_{SW}$ & 7.2$\times 10^{16}$ cm & 3 \\
T$_{in}$ & 150 K & 3 \\
F$_{T}$ & 0.4\rlap{3} & 3 \\
$\epsilon$ & 2.0 & 3 \\
$\epsilon_2$ & 0.2\rlap{5} & 3 \\
\.M$_{SW}$ & $3\times 10^{-5}$ M$_\odot$ yr$^{-1}$ & 3 \\
\.M$_{AGB}$ & $5.1\times 10^{-6}$ M$_\odot$ yr$^{-1}$ & 3 \\
\.M$_{AGB}$/\.M$_{SW}$ & 0.1\rlap{7} & 3 \\
M$_{gas}$  &  0.059 M$_\odot$ & 3 \\ 
t$_{dyn}$ & 1240 yr & 3 \\
SW duration & \phn840 yr & 3 \\
AGB duration & 6570 yr & 3 \\
\enddata
\tablerefs{(1) \cite{hony2003} (2) \cite{vanwinckel2000} (3) this work
   }
\end{deluxetable}

\begin{deluxetable}{lcc}
\tablewidth{0pt}
\tablecaption{Model Parameters  of HD~56126:  Dust emission\label{modeldusttab}}
\tablehead{
\colhead{Parameter} &\colhead{Value} & \colhead{Ref.\ and Notes}}%$^1$\\
\startdata
\textit{Input:} \\
D  & 2.4 kpc & 1 \\
Spec Type & F1-2I & 2 \\
T$_{eff}$ & 7250 K & 2 \\
L$_*$ & 6090 L$_\odot$ & 1 \\
R$_*$ & 3.45$\times 10^{12}$ cm & 3 \\
R$_{in}$ & 4.3$\times 10^{16}$ cm & 3 \\
R$_{SW}$ & 7.2$\times 10^{16}$ cm & 3 \\
R$_{out}$ & 3.0$\times 10^{17}$ cm & 3 \\
V$_{exp}$ & 10.5$\pm 0.5$ km s$^{-1}$  & 3\\
ISM A$_V$ & 0.5 & 4 \\
t$_{dyn}$ & 1240 yr & 3 \\
SW duration & \phn840 yr & 3 \\
AGB duration & 6570 yr & 3 \\
\\
\textit{Output:}  \\
$\tau_{eq} (10.3)$ &  0.025 & 3 \\
T$_{in}$(dust) & 150 K & 3  \\
a$_{min}$, a$_{max}$ & 0.001, 0.01 $\mu$m & 3,5 \\
A,B,C,D,E,F & 3,2,2.5,3,3,1.5 & 3,6 \\
\.M$_{equator}$/\.M$_{pole}$ & 4 & 3 \\
inclination angle & 80\arcdeg & 3 \\
\.M$_{SW}$(dust) & $1.9\times 10^{-7}$ M$_\odot$ yr$^{-1}$ & 3 \\
\.M$_{AGB}$(dust) & $9.6\times 10^{-8}$ M$_\odot$ yr$^{-1}$ & 3 \\
\.M$_{AGB}$/\.M$_{SW}$(dust) & 0.5 & 3 \\
M$_{dust}$ & $7.8\times 10^{-4} $ M$_\odot$ & 3 \\ 
M$_{gas}$/M$_{dust}$ & 75 & 3\\
\enddata
\tablerefs{(1) \cite{hony2003} (2) \cite{vanwinckel2000}  (3)~this work
 (4)~\cite{meixner1997} (5)~We adopt the size distribution function of the form
$n(a)= a^{-\gamma} e^{-a/a_{\rm max}}$ for $a > a_{\rm min}$
\citep{kim94}. (6)~A-F are parameters in the density function defined in equation 1 of \cite{ueta2003}. }
\end{deluxetable}

\begin{figure}
\caption {\textbf{Figure 1 is on the following page.}  
HD 56126 CO J=1-0 line emission.  \textit{Top}: Channel maps  of the combined BIMA B,C, and D 
array data. The velocity  width of each channel is 2 km s$^{-1}$ and the central velocity in km s$^{-1}$ 
is located in the top right corner of  each channel map. The contour levels are 
 $-$0.5, $-$0.25, 0.25, 0.5, 0.75, 1.0, 1.25, 1.5, 1.75, and 2 Jy/beam. 
The FWHM beam size is located in the bottom
right corner of the first channel map. A cross marks the location of the central star which is 0.\arcsec 63
to the west of the map's phase center. The horizontal
line in the first channel map provide a physical size, 4800 AU, of 2\arcsec. \textit{Middle}:  The best
fit model channel maps of the CO J=1-0 line  using the same contour spacing and grey scale as the data channel
maps. All other markings are as shown in the data channel maps. 
\textit{Bottom Left}: The CO J=1-0 total flux line profile: the dashed line shows the data and
the solid line shows the model.  \textit{Bottom Right}: The azimuthally averaged radial profile of the
73 km s$^{-1}$ channel map.  The solid line is the model and the points with error bars are the data.
Note that the error bars represent the variation in the intensity due to the clumpy nature of the CO
emission.\label{cofig} }
\end{figure}

\addtocounter{figure}{-1}
\begin{figure}
\plotone{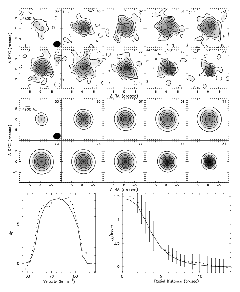}
\caption{ }
\end{figure}

\begin{figure}
\plotone{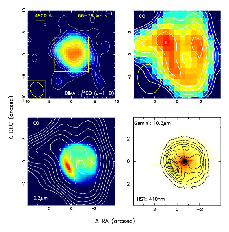}
\caption{Comparison of the  CO emission (BIMA) with the tracers of dust: mid-infrared emission (Gemini) and 
scattered optical light (HST). \textit{Top Left}:  CO emission from the combined B, C and D arrays
is averaged over the velocity range 66 to 78 \kms\ and shown with
 contour levels  of 0.12 Jy/beam (10\% of peak). 
The beam size of 3\farcs5$\times$2\farcs9 is shown
in the bottom left corner.
The yellow square in the center of the image corresponds to the area covered in the 
other three images.
\textit{Top Right}:  B-array only data is averaged over the velocity range  66 to 78 \kms\ 
and shown with contour levels
of 0.05 Jy/beam.  The beam size of  2\farcs6$\times$2\farcs0 is shown
in the bottom left corner. \textit{Bottom Left}:
The B-array contours compared to the mid-IR image shown as a colored image. Note
that the central star is located in the central cavity of CO emission
and both are offset 0\farcs63 west of the BIMA maps phase center.
\textit{Bottom Right}: The Gemini 10.3 $\mu$m image
\citep{kwok2002} shown as contontours with levels are 10\% of peak overlayed on
the  HST optical image shown as a colored image \citep{ueta2000}.
 \label{comidirhst}}
\end{figure}

\begin{figure}
\epsscale{.6}
\plotone{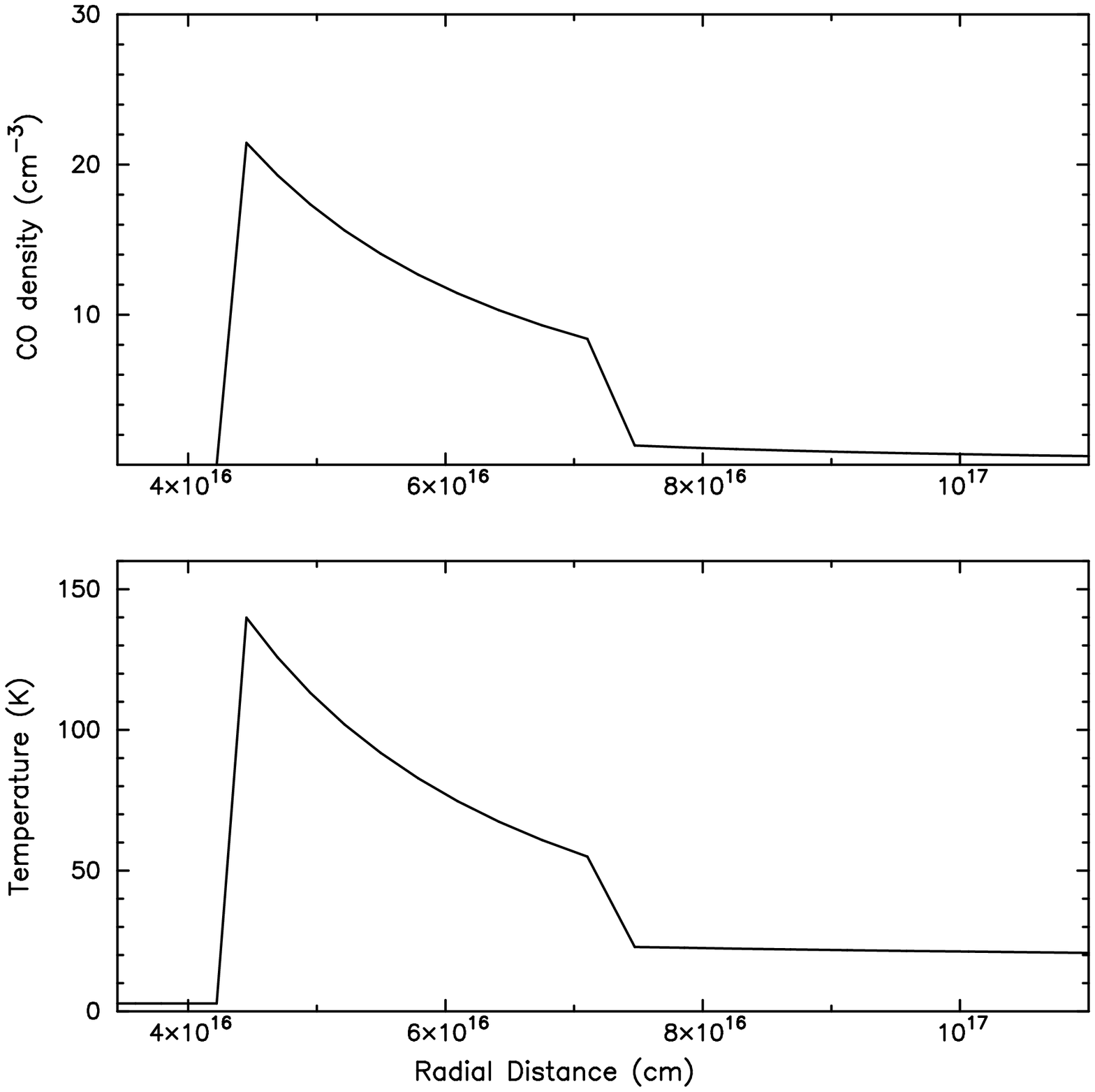}
\caption{ The density (top) and temperature (bottom)  of the molecular gas versus the radius in the
circumstellar envelope. The central star is located at 3.45$\times 10^{12}$ cm.  
These profiles serve as inputs to the CO
modeling process.  \label{dentempfig}}
\end{figure}

\begin{figure}
\epsscale{.75}
\plotone{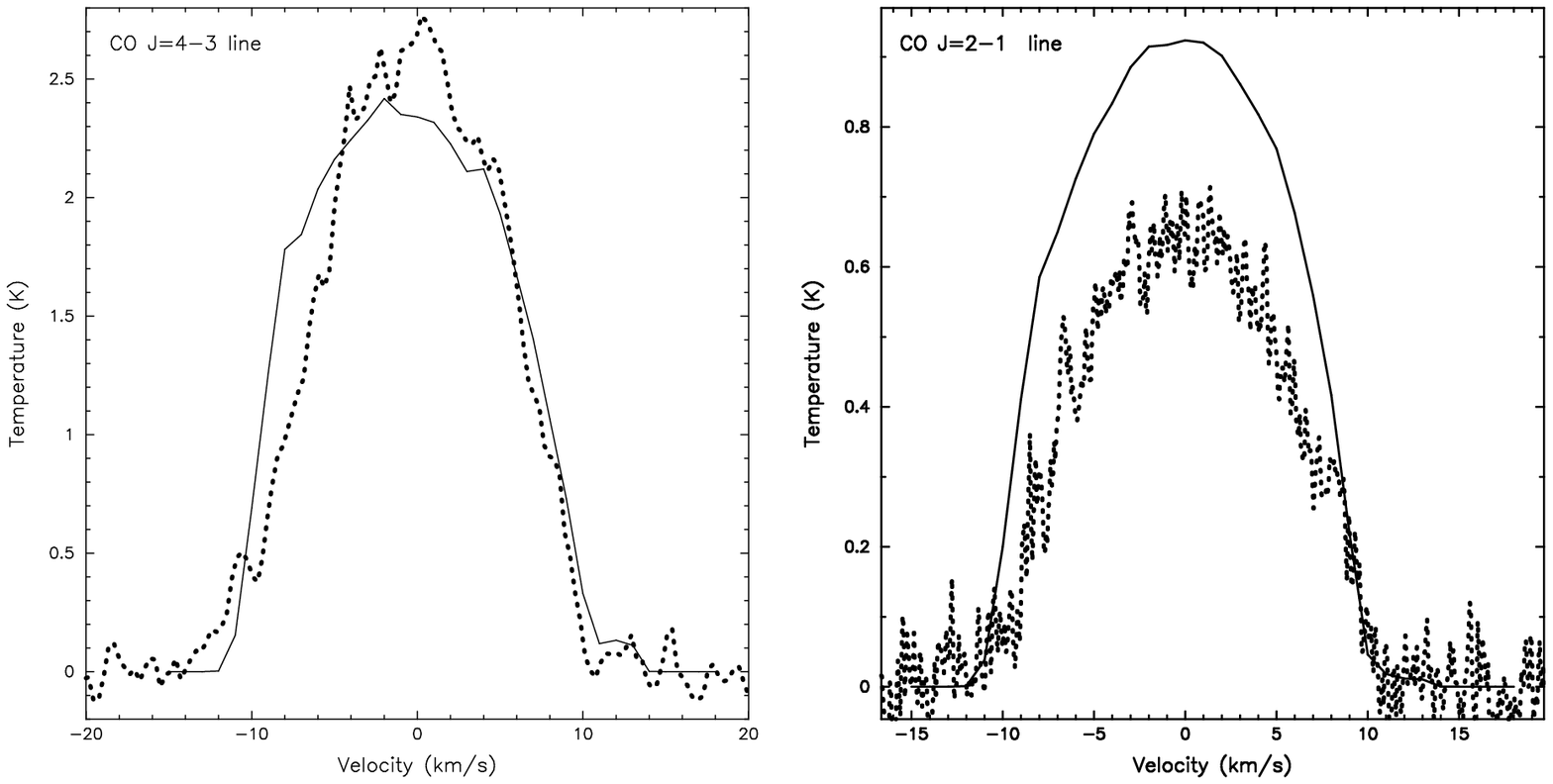}
\caption{Comparison of the CO model results (solid line) with the data  (dashed lines)
for the CO J=4-3 line (left; \citep{knapp2000}) and for the CO J=2-1 line (right; \citep{knapp1998}).  
The velocity scale is relative  to the systemic velocity.
\label{linecofig}}
\end{figure}

\begin{figure}
\epsscale{.9}
\plotone{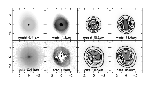}
\caption{Comparison of model images from our dust modeling (top) with images from the literature
(bottom): 410 nm from \cite{ueta2000}, 1.2 $\mu$m  from \cite{ueta2004}, 10.3 and 18.0 $\mu$m
from \cite{kwok2002}. Grey scale ranges for model and image are the same.
All contours are 10\% of peak contours. \label{moddustfig}}
\end{figure}

\begin{figure}
\epsscale{.6}
\plotone{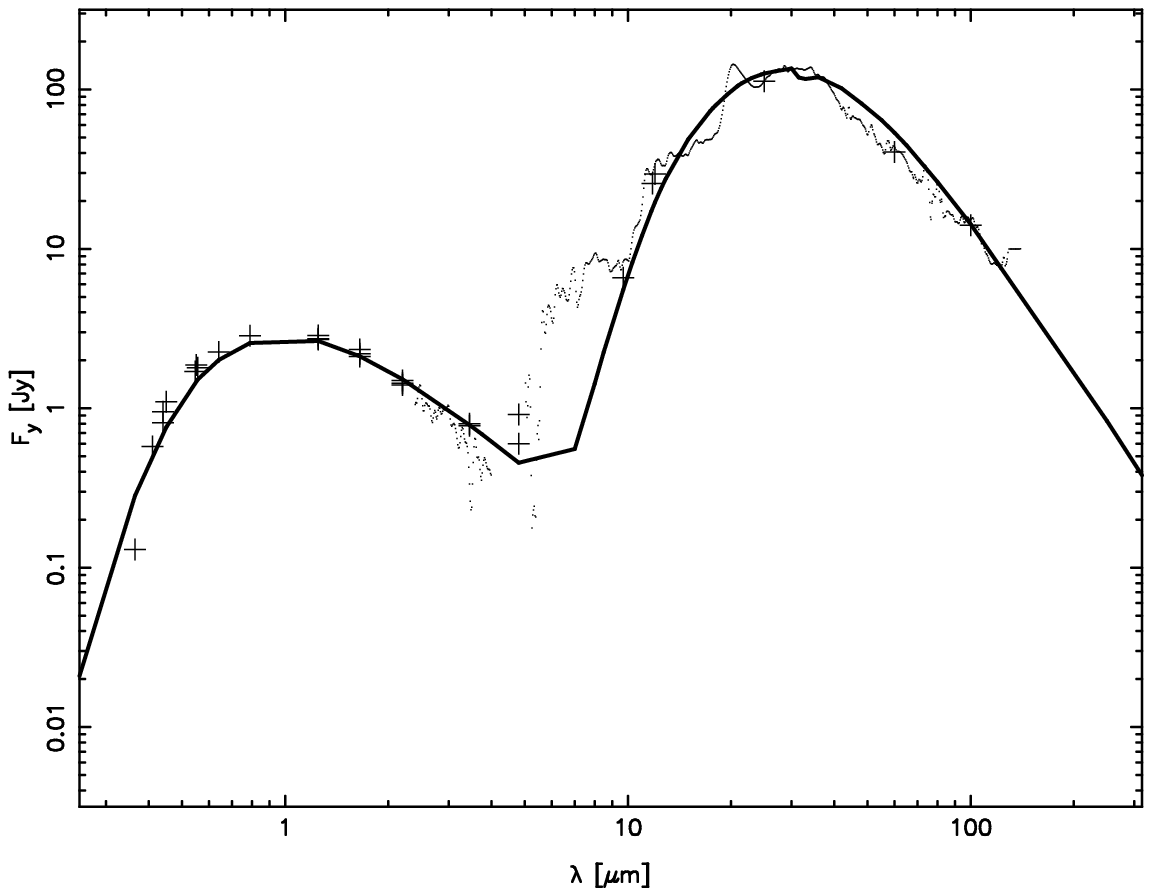}
\caption{The spectral energy distribution of HD 56126. The dust model shown as a solid line. The
photometry shown as crosses from \cite{hrivnak1989}  and \cite{hony2003}. 
The ISO spectrum from \cite{hony2003} shown as dots. \label{sedmodfig}}
\end{figure}

\end{document}